\documentclass[12pt]{spieman}  
\usepackage{amsmath,amsfonts,amssymb}
\usepackage{graphicx}
\usepackage{setspace}
\usepackage{tocloft}
\usepackage{xcolor}

\newcommand{\degree}{$^{\circ}$}

\title{The Complex Magnetic Field of the Extreme Galactic Center: PRIMA Science Potential}

\author[a,*]{Dylan M. Par\'e}
\author[a]{David T. Chuss}
\author[b,c]{Kaitlyn Karpovich}
\author[d]{Natalie Butterfield}
\author[e]{Edward J. Wollack}
\author[f]{Mark R. Morris}
\author[a]{Jeffrey Inara Iuliano}
\affil[a]{Villanova University, Department of Physics, 800 E. Lancaster Ave., Villanova, USA, 19085}
\affil[b]{Stanford University, Department of Physics, Stanford, USA, 94305}
\affil[c]{Kavli Institute for Particle Astrophysics \& Cosmology, P.O. Box 2450, Stanford, CA, USA, 94305}
\affil[d]{National Radio Astronomy Observatory, 520 Edgemont Road, Charlottesville, VA, USA}
\affil[e]{NASA Goddard Spaceflight Center, 8800 Greenbelt Rd, Greenbelt, MD 20771, USA}
\affil[f]{University of California, Los Angeles, Department of Physics \& Astronomy, 475 Portola Pl., Los Angeles, CA 90095-1547, USA}

\cftpagenumbersoff{figure}
\cftpagenumbersoff{table} 
\begin{document} 
\maketitle

\begin{abstract}
The Central Molecular Zone (CMZ) of the Galactic Center (GC) region of the Milky Way contains a substantial fraction of the molecular mass of the Galaxy ($\geq$10$^{7}$ M$_{\odot}$) yet exhibits an order of magnitude lower star formation efficiency (SFE) than expected given the high densities found in this region. There are multiple possible explanations for the depressed SFE in the CMZ, like feedback, strong turbulence, longer free-fall timescales, and high magnetic field strengths. It is currently unclear which of these mechanisms is the dominant inhibitor of star formation in the CMZ. It is important to understand the star formation process in the extreme environment of the CMZ because it is the only Galactic nuclear region we are able to study at high spatial resolutions with current observatories. One way to determine the relative importance of the different SFE inhibiting mechanisms is through multi-spatial and multi-frequency polarimetric observations of the CMZ. Such observations will provide insight into the behavior of the magnetic field in this unique environment. These observations will complement radio observations of non-thermal structures revealing the magnetic field morphology and polarization.  The PRobe far--Infrared Mission for Astrophysics (PRIMA) will be uniquely capable of contributing to such explorations by providing unique resolutions and frequencies for polarimetric observations. The PRIMAger instrument will yield polarimetric observations covering the wavelength range 80 -- 261 $\mu$m with beam sizes ranging from 11 -- 28$^{\prime\prime}$, capabilities that complement existing and upcoming observatories.

\end{abstract}


{\noindent \footnotesize\textbf{*}Dylan M. Par\'e,  \linkable{dylanpare@gmail.com} }

\begin{spacing}{2}   

\section{Introduction}
\label{sect:intro}  

\begin{figure}
    \begin{center}
        \begin{tabular}{c}
        \includegraphics[width=1.0\textwidth]{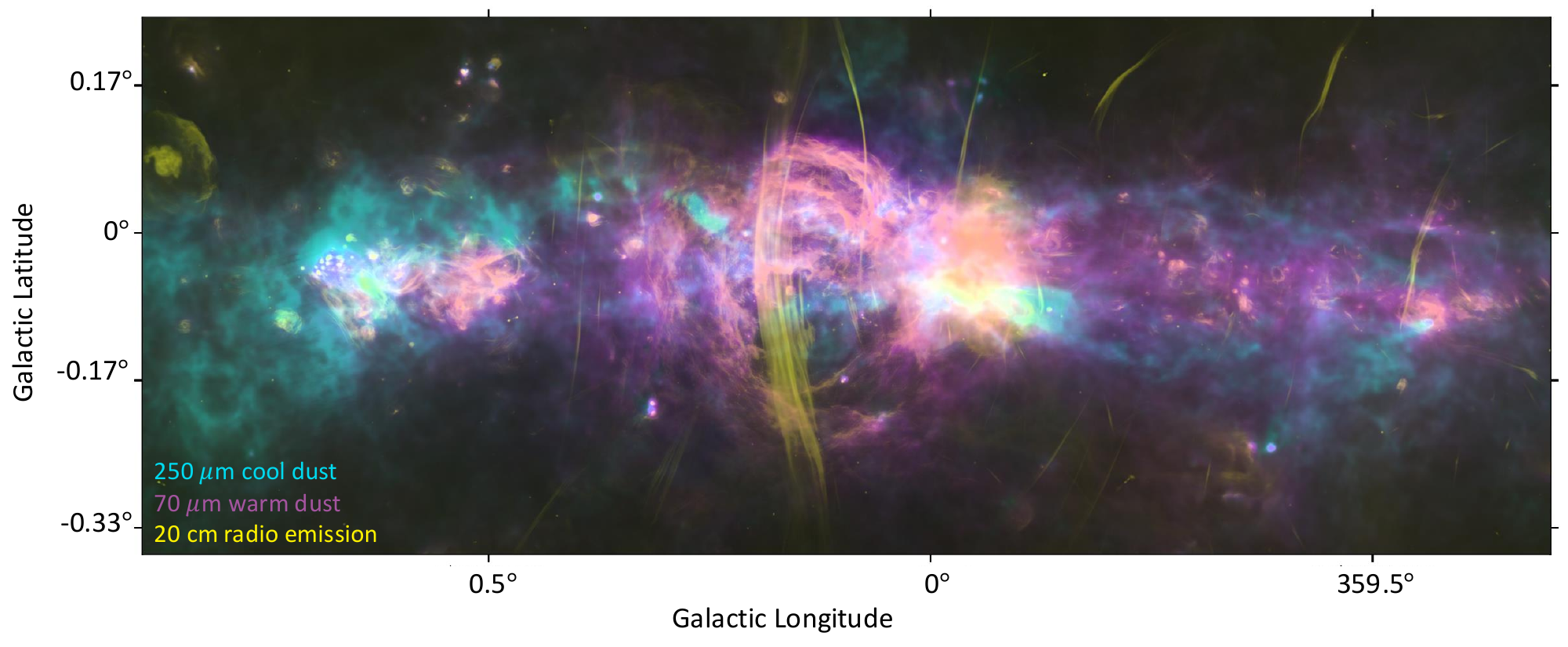}
        \end{tabular}
    \end{center}
    \caption
    { \label{fig:legend}
A 3-color view of the CMZ\cite{Pare2024} with 20 cm (1.28 GHz) MeerKAT radio emission in yellow and 250 $\mu$m cool dust and 70 $\mu$m warm dust emission in cyan and magenta, respectively, both observed by Herschel\cite{Molinari2011,Heywood2022}.}
\end{figure}

The central region of the Milky Way, known as the Central Molecular Zone (CMZ), contains 4\% (or $\geq$10$^{7}$ M$_{\odot}$) of the molecular mass of the Milky Way\cite{Barnes2017}. The CMZ consists of numerous high density clouds (with densities ranging from 10$\rm^3$ -- 10$\rm^6$ cm$^{-3}$), elevated turbulent linewidths (10 -- 50 km s$^{-1}$), and strong magnetic field strengths of 100s of $\mu$G even in low density portions of the CMZ\cite{Pillai2015,Mills2018,Hsieh2018,Henshaw2019}. Figure \ref{fig:legend} displays the CMZ as probed by infrared and radio observations revealing the complex thermal and non-thermal structures observed in the region\cite{Pare2024}. Many of the non-thermal features are organized as filamentary structures collectively known as the non-thermal filaments (NTFs). The NTFs shown in the radio emission in Figure \ref{fig:legend} are unique to the GC, and imply the existence of a CMZ-wide vertical magnetic field oriented perpendicular to the Galactic disk. The magnetic field inferred from far-infrared dust polarization is, conversely, observed to be a horizontal field oriented parallel to the Galactic disk\cite{Guan2021}. The complexity of the magnetic field in this region indicates that the magnetic field in this region may be of particular dynamical importance.

The extreme conditions of the CMZ, coupled with its relative proximity to Earth (being only $\sim$8 kpc away \cite{Abuter2019}) make it an important local analog to more distant galactic nuclear regions. Observations of the CMZ can therefore provide insight into structures like the individual molecular clouds present within a galactic nuclear region. The star formation efficiency (SFE) in the CMZ is only 10\% of that which would be expected given the high densities found there (predicted star formation rate of $\sim$0.5 M$_{\odot}$ yr$^{-1}$ compared to an actual star formation rate of $\sim$0.02 M$_{\odot}$ yr$^{-1}$)\cite{Longmore2013,Mills2018,Morris2023}. There are multiple possible explanations for the low SFE in the CMZ, which include the high strength of the magnetic field, high velocity gradients caused by strong tidal forces, plasma compressibility as a result of the high field strengths, and elevated turbulence\cite{Krumholz2015}. Understanding how each of these factors is inhibiting star formation in the GC will expand our understanding of the properties of star formation in extreme environments like the CMZ and other galactic nuclear regions.

The magnetic field in the molecular component in this region can be probed at sub-millimeter and far-infrared wavelengths using dust polarization observations. Dust grains in the presence of a magnetic field are preferentially aligned with their long axis perpendicular to the orientation of the field, resulting in the thermal emission of light polarized with an orientation perpendicular to the plane-of-sky magnetic field direction. The favored mechanism is known as radiative torque (RAT) alignment\cite{Lazarian07,Anderson2015}. In the proceeding discussion and throughout the paper we assume that the dust grains in the CMZ are aligned by this mechanism.

\begin{figure}
    \begin{center}
        \begin{tabular}{c}
        \includegraphics[width=1.0\textwidth]{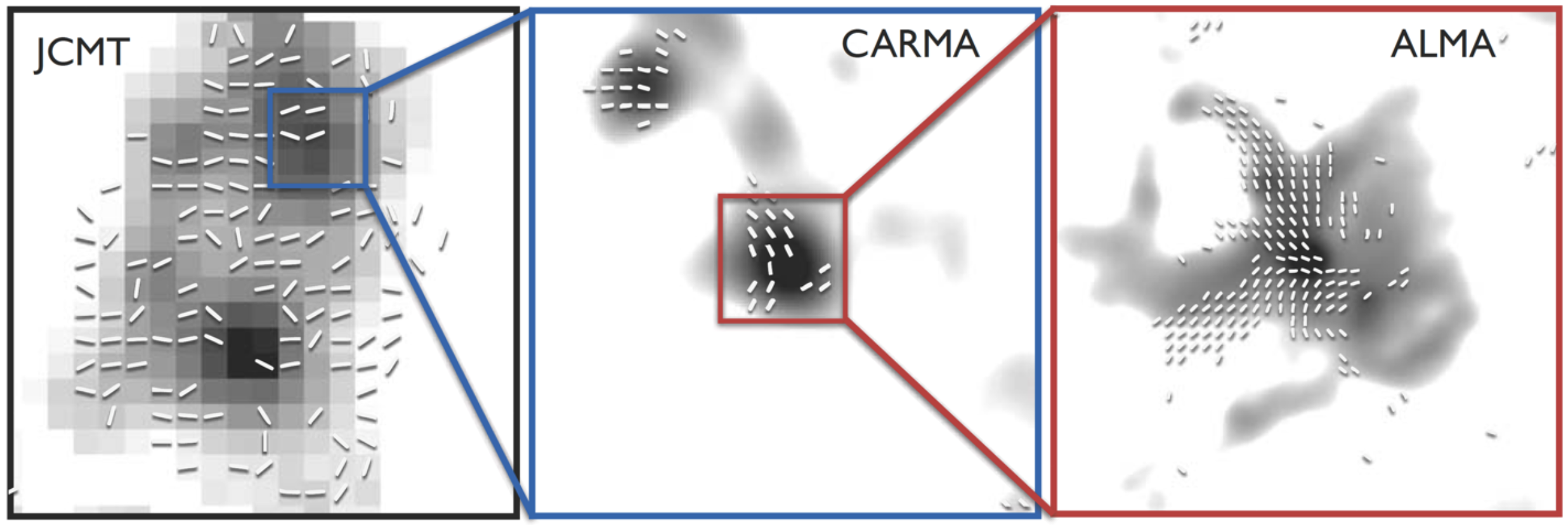}
        \end{tabular}
    \end{center}
    \caption
    { \label{fig:multi}
As an example of a multi-scale magnetic field analysis, we show the (non-GC) star-forming region Ser-emb-8 \cite{Hull2017}. The James Clerk Maxwell Telescope (JCMT) image has a spatial resolution of $\rm \sim$10,000 AU, the CARMA image has a spatial resolution of $\rm\sim$1000 AU, and the Atacama Large Millimeter/submillimter Array (ALMA) image has a spatial resolution of $\rm\sim$100 AU.}
\end{figure}

One way to assess the impact of the magnetic field strength and turbulence on star formation is to study the structure of magnetic fields in different physical environments over multiple spatial scales ranging from, for example, protostellar core scales (1000s of AU, or $\sim$5 mpc) to molecular cloud scales (100,000 AU, or $\sim$ 0.5 pc)\cite{Zhang2014}. Such multi-scale analyses have been conducted in the Galactic disk for numerous star-forming regions\cite{Liu2022}. For example, the magnetic field of the star forming region Ser-emb-8 was studied over a range of spatial scales from 100 AU to 10,000 AU (5$\times$10$^{-4}$ to 5$\times$10$^{-2}$ pc) and it was determined that the turbulence of the cloud was more significant in inhibiting star formation than the magnetic field\cite{Hull2017}. The multi-spatial observations applied to that source are shown in Figure \ref{fig:multi}. The conclusion that turbulence dominates over the magnetic field for this star forming region was reached because of the disconnect observed between the magnetic field configuration encountered over a broad range of spatial scales. Such a disconnect was also predicted from MHD simulations of this source, which further motivated the conclusion that turbulence dominates in this cloud. This kind of multi-scale magnetic field analysis (in conjunction with MHD simulations) has not been conducted in the CMZ, even though the CMZ possesses a significant reservoir of dense molecular dust.

In addition to a multi-scale analysis, the strength of the magnetic field can be estimated using the Davis-Chandrasekhar-Fermi (DCF) method.\cite{Davis1951,CF1953}. The combination of a multi-scale imaging analysis of the magnetic field morphology and studies of spectral lines probing turbulence will enable a comprehensive assessment of the dynamical importance of the magnetic field. Studying the geometry of the magnetic field has other important utility for deepening our understanding of the role of the magnetic field in the CMZ. The observations of the magnetic field can be compared to MHD models to determine the importance of turbulence in the CMZ clouds.\cite{Reissl2016,Hull2020}.

PRIMA observations will provide a key range of resolutions for a multi-scale polarimetric analysis of star-forming regions in the GC. Furthermore, the PRIMA FIRESS spectrometer can be used in conjunction with imaging obtained from PRIMAger to enable an analysis of spectral lines tracing the turbulence within CMZ clouds. A good candidate line is the hydrogen deuteride line at 112 $\mu$m\cite{Wright1999,Breysse2022}. FIRESS  will therefore be able to provide the information needed to estimate magnetic field strengths using DCF.

In this work we detail possible future strategies for deepening our understanding of the CMZ magnetic field using the PRobe far-Infrared Mission for Astrophysics (PRIMA) observatory. In Section \ref{sec:prev} we discuss the current state of our understanding regarding previous observations of the magnetic field in the CMZ. In Section \ref{sec:prima} we present the exciting science potential of PRIMA that will enable even more powerful studies of CMZ magnetic fields. In Section \ref{sec:syn} we discuss synergy between these aspirational observations and other current and next-generation instruments.

\section{Previous CMZ Polarimetric Observations} \label{sec:prev}

\begin{figure}
    \begin{center}
    \begin{tabular}{c}
    \includegraphics[width=0.95\textwidth]{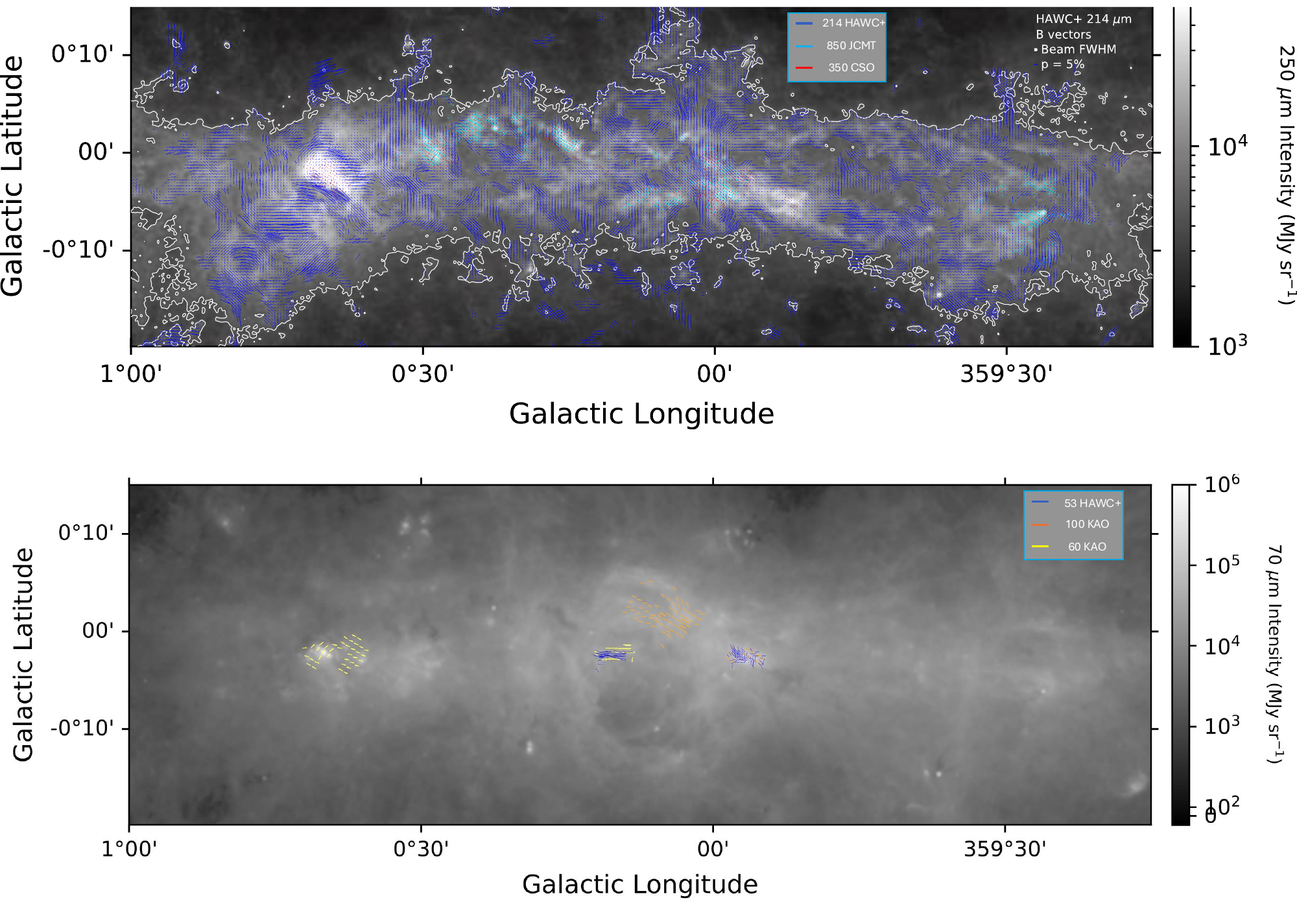}
    \end{tabular}
    \end{center}
    \caption 
    { \label{fig:2-color}
Comparison of the magnetic field observed towards the CMZ for the cool (top) and warm (bottom) dust components, where the dust temperature ranges from 15 -- 40 K\cite{Molinari2011}. Top panel: the magnetic field inferred from polarimetric measurements dominated by cool dust. The grayscale background is 250 $\mu$m Herschel emission of the CMZ \cite{Molinari2011}. Overlaid colored dashes indicate magnetic field orientations derived from SOFIA/HAWC+ 214 $\mu$m (blue), CSO 350 $\mu$m (red), and JCMT 850 $\mu$m (cyan) \cite{Dotson2010,Lu2023,Pare2024}. Bottom panel: the magnetic field inferred from observations of warm dust. The grayscale background is 70 $\mu$m Herschel emission from the CMZ \cite{Molinari2011}. Overlaid colored dashes indicate magnetic field orientations derived from SOFIA/HAWC+ 53 $\mu$m (blue), and KAO 100 and 60 $\mu$m observations (yellow and orange) \cite{Dotson2000,Dotson2010,Guerra2023}.} 
\end{figure} 

The magnetic field towards the CMZ, as observed at far-infrared wavelengths, is complex. At arcminute resolutions (angular resolution of 1 -- 2$'$, physical resolution of $\sim$10$^6$ AU or 3.58 pc) using the PILOT and ACTpol surveys, the magnetic field is generally ordered and is largely parallel to the Galactic plane\cite{Mangilli2019,Guan2021}. One possibility for this ordered, parallel field could be dust that is foreground to the central 150 pc of the Galaxy\cite{Pare2024}. Higher-resolution observations at an angular resolution of $\sim$19.6$^{\prime\prime}$ (corresponding to a physical scale of $\sim$10$^5$ AU or 0.78 pc) were recently obtained from the Stratospheric Observatory for Infrared Astrophysics (SOFIA) using the High-resolution Airborne Wideband Camera Plus (HAWC+) instrument at 214 $\mu$m \cite{Butterfield2024a,Butterfield2024b,Pare2024}. The magnetic field obtained from these SOFIA/HAWC+ observations is shown in the upper panel of Figure \ref{fig:2-color}. The blue dashes show the magnetic field inferred from the 214 $\mu$m SOFIA/HAWC+ observations, with red and cyan dashes representing other submillimeter observations made using the Caltech Submillimeter Observatory (CSO)\cite{Dotson2000} and the James Clerk Maxwell Telescope (JCMT) SCUBA-2/POL-2\cite{Dotson2010,Lu2023} instrument, respectively. These observations show the magnetic field that threads the cool dust in the CMZ. The magnetic field revealed by these observations is more complicated than that inferred from PILOT and ACTpol with lower angular resolution, possibly indicating that FIREPLACE is largely probing the magnetic field local to the CMZ rather than some other foreground field component\cite{Pare2024}.

In contrast to the magnetic field observed in the far-infrared, the magnetic field observed at radio wavelengths using instruments like the Very Large Array (VLA) reveals a largely vertical (perpendicular to the Galactic plane) geometry\cite{Lang1999a,Lang1999b,LaRosa2004}. It remains unclear how the radio and far-infrared magnetic fields connect. Although the SOFIA/HAWC+ observations have unveiled the magnetic field structure in the CMZ in most of the FIR bright cloud regions, there is considerable room for further improvement. In particular, the SOFIA/HAWC+ observations lack sensitivity to fields on scales larger than the 19.6$^{\prime\prime}$ ($\sim$0.7 pc) beam size of SOFIA/HAWC+, meaning that the magnetic field distribution in fainter regions of the CMZ are not well determined\cite{Pare2024}. PRIMA will enable an investigation of the large-scale fields that SOFIA/HAWC+ was not sensitive to. The extremely low background provided by PRIMA enables improved sensitivity to fainter CMZ structures. Investigating these fainter regions is important for probing how the magnetic field in the CMZ clouds connects to the vertical field system traced by the GC NTFs. Studying this connection will help determine how the seemingly distinct vertical and horizontal magnetic fields in the CMZ evolved. Studying such connections (or the lack thereof) can refine our understanding of the line-of-sight geometry of the thermal dust clouds and the NTFs. For example, if there are no connections between the magnetic fields traced by the NTFs and thermal dust it could indicate these structures are located at different points along the line-of-sight and are not interacting.

In addition, the improved sensitivity of PRIMA will enable the  study of high-velocity gas located at the extreme edge of the CMZ. A recent study revealed extended cloud structures at high velocities (100s km s$^{-1}$) that could be a result of turbulent, shocked gas overshooting the CMZ after brushing by it\cite{Veena2024}. These are faint, diffuse structures. meaning that the improved sensitivity and larger beam size of PRIMA compared to SOFIA/HAWC+ will be much better suited to studying the polarization of these structures. This high velocity gas also connects to the bar and dust lanes of the Galaxy\cite{Yang2024}, which the high velocity gas could be flowing along. PRIMA could therefore refine our understanding of these high-velocity structures by allowing us to probe the magnetic fields local to these clouds and how they compare to the higher density CMZ clouds. This understanding will be achieved by comparing how the polarization and magnetic field orientation varies in different cloud morphologies\cite{Pare2024}. It will also be possible to compare how the magnetic field aligns with the molecular structure of the clouds using statistical methods like the Projected Rayleigh Statistic (PRS)\cite{Jow2018}, which will make it possible to assess whether the conditions in these clouds parallel those observed in CMZ clouds or in Galactic disk clouds \cite{Pare2025,PlanckXXXV}. This understanding of high-velocity structures will make  it possible to assess impact of the magnetic field on the dynamics of the CMZ gas and connections between the CMZ and Galactic dust lanes.

PRIMA also extends to wavelengths down to $\sim$90 $\mu$m, which are crucial for two reasons. First, Herschel observations have revealed that warm dust in the Galactic center has a morphology distinct from that of the cooler dust that dominates at longer wavelengths (see the lower panel of Fig.~\ref{fig:2-color}). Polarimetry at shorter wavelengths probes the field in regions where energetic events have heated the grains. The magnetic field geometry in these regions may provide insight into the role and response of the fields in such events. The magnetic field measurements at these locations provide a means to probe possible connections between the fields across a broad range of temperatures, column densities, and energy injection rates.
\begin{figure}
    \begin{center}
    \begin{tabular}{c}
    \includegraphics[width=1.0\textwidth]{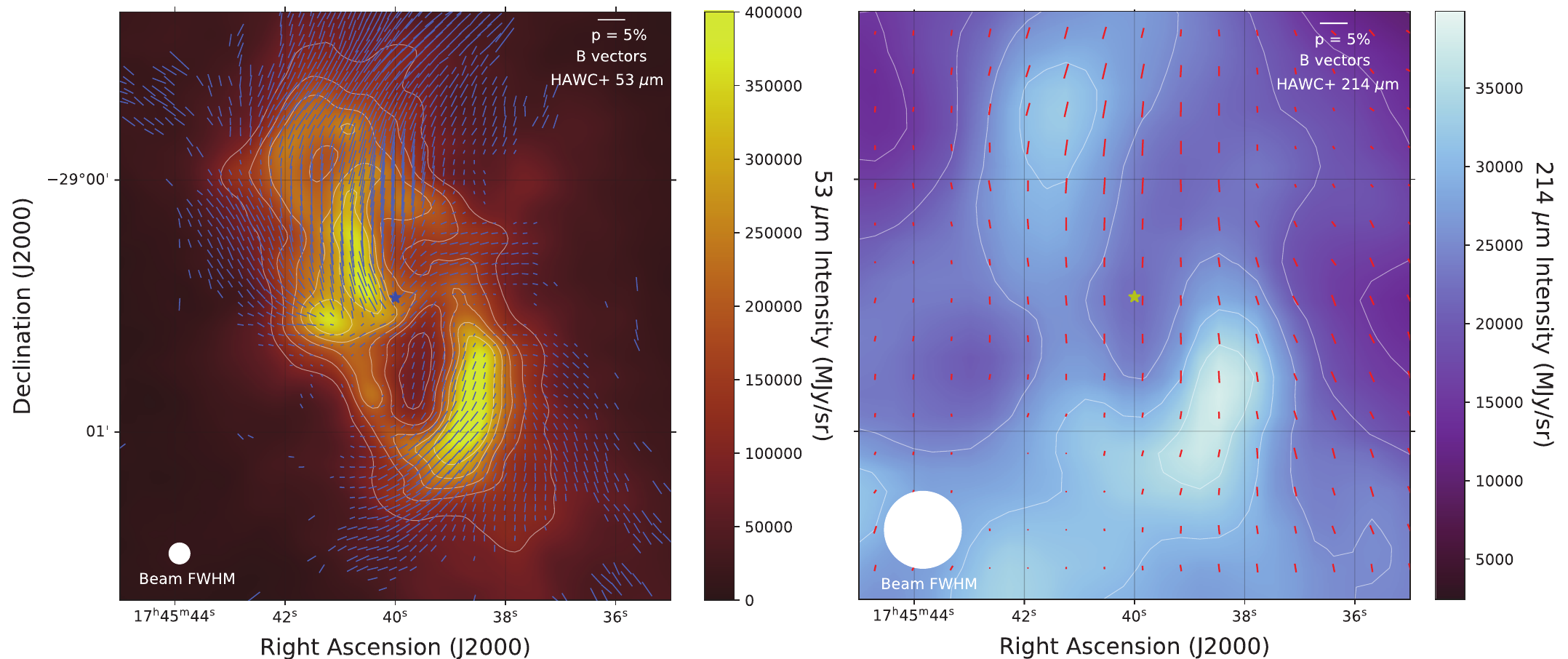}
    \end{tabular}
    \end{center}
    \caption 
    { \label{fig:CND}
    The Galactic center's circumnuclear ring has been observed by HAWC+/SOFIA at 53 and 214 $\mu$m in both total intensity and polarization\cite{Guerra2023,Harper2018}. The morphology and magnetic field geometry differ significantly at the two wavelengths. This is likely due to superposition of multiple cloud components along the line of sight with distinct temperatures and dust densities. Multi-wavelength polarimetric imaging using PRIMA can be employed to elucidate this superposition, better understand the geometry of the dust, and untangle the interaction of the magnetic field with the strong shear of the CND.}
\end{figure}
The lower panel of Figure \ref{fig:2-color} shows the 70 $\mu$m warm dust probed by Herschel with the existing polarimetric observations from $\sim$50 -- 100 $\mu$m using the Kuiper Airborne Observatory (KAO). We do not currently have much insight into the magnetic field in regions containing warm dust despite the fact that it is a morphologically distinct dust population. Furthermore, there is currently no way to study the magnetic field of the warm dust at sub-arcminute resolutions.

Given the different dust populations, the sightlines in the Galactic center are complicated. For example, Figure~\ref{fig:CND} shows observations of the circumnuclear (CND) disk at 53 and 214 $\mu$m from SOFIA/HAWC+ \cite{Guerra2023, Pare2024}. The CND is a cloud orbiting the central black hole of the Galaxy, Sgr A$^*$, placing it at the heart of the CMZ\cite{Hsieh2017}. The CND is an important reservoir of magnetized gas and dust that is likely facilitating accretion onto the Galactic black hole\cite{Solanki2023,Blank2016}. Furthermore, the magnetic field of the CND is strongly interacting with and being enhanced by the strong shear of the orbiting gas\cite{Wardle1990,Guerra2023}. Furthermore, there is evidence that the polarization observed at different wavelengths for the CND is tracing different dust layers along the line of sight, such as a cool outer dust cloud and a warmer structure within the inner 5 pc of the GC \cite{Akshaya2024}. The key to understanding such complex environments is multiwavelength total intensity and polarimetric observations. PRIMA uniquely covers the short wavelengths that are crucial for constraining the radiative and magnetic environments of the GC clouds. 

\section{Utility of PRIMA for Studies of the Galactic Center} \label{sec:prima}
\begin{table}[ht]
\caption{Observational Parameters for PRIMA GC Map} 
\label{tab:fonts}
\begin{center}       
\begin{tabular}{|l|l|} 
\hline
\rule[-1ex]{0pt}{3.5ex} Observation Parameter & Values$^i$  \\
\hline\hline
\rule[-1ex]{0pt}{3.5ex}  Map Size & $\sim$1.5\degree$\times$0.5\degree \\
\hline
\rule[-1ex]{0pt}{3.5ex}  PRIMA Instrument & PRIMAger   \\
\hline
\rule[-1ex]{0pt}{3.5ex}  Wavelength Used & 96, 126, 172, 235 $\mu$m   \\
\hline
\rule[-1ex]{0pt}{3.5ex}  Beam FWHM & 11, 15, 20, 28$^{\prime\prime}$  \\
\hline
\rule[-1ex]{0pt}{3.5ex}  Target Total Power Sensitivity &  386, 505, 425, 475 MJy sr$^{-1}$ \\
\hline
\rule[-1ex]{0pt}{3.5ex} Target Polarization Sensitivity & 3.86, 5.05, 4.25, 4.75 MJy sr$^{-1}$ \\
\hline
\rule[-1ex]{0pt}{3.5ex} Time Required to Reach Sensitivity & 3.3, 1.1, 0.9, 0.4 hrs  \\
\hline 
\end{tabular}
\end{center}
\footnotesize{$^i$ PRIMA values are taken from https://prima.ipac.caltech.edu/ and Ciesla+2025 which is also part of this Jatis special issue.}
\label{tab:PRIMA}
\end{table}
A key method for assessing the importance of the magnetic field in a star-forming region is the study of how the magnetic field distribution varies over a range of spatial scales (like in the example shown in Fig. \ref{fig:multi}). PRIMA will provide an important spatial scale for this kind of multi-scale analysis, probing the molecular cloud scale ($\sim$10$^5$ AU). Although this is the same scale probed by SOFIA/HAWC+, the higher sensitivity of the PRIMA observations will make it possible to conduct this analysis in fainter, more diffuse regions of the CMZ. The magnetic field derived from the PRIMA observations can then be compared to that derived from ALMA, which will have a much higher resolution sensitive to protostellar envelope spatial scales. For example, an ALMA proposal in Cycle 11 to observe the largest molecular cloud in the CMZ was approved for time and will likely get observation time (proposal ID: 2024.1.00140.S, PI: Q. Zhang). This multi-scale comparison is where a next-generation instrument with high sensitivity can enable a similar analysis without requiring an impractical amount of observation time.

PRIMA will be uniquely suited to furthering polarimetric studies of the magnetic field in the CMZ covering a range of wavelengths. Such studies will enhance our understanding of the impact of the magnetic field on the process of star formation, a fundamental process in astronomy. Furthermore, with the discontinuation of SOFIA, PRIMA will be the only far-infrared instrument capable of making key polarimetric measurements of the magnetic field in the warm dust in the CMZ. The warm dust appears to be a distinct component of the CMZ as seen in Figure \ref{fig:2-color} and \ref{fig:CND}. We note that since the FIREPLACE observations are largely observed to trace the CMZ clouds rather than some foreground component, the PRIMA observations of bright CMZ clouds will also be dominated by the CMZ emission. For fainter regions there may be more significant contamination by other clouds along the line-of-sight. For these fainter regions the potential impact of line-of-sight contamination will be assessed by studying the total and polarized intensity morphologies with the magnetic field orientations to assess whether they are tracing the same structure. If so, this indicates that the emission is dominated by the CMZ cloud contribution. If not, it would indicate that significant foreground contamination is occurring. This was the analysis employed by the FIREPLACE collaboration to quantify the extent of line-of-sight contamination in their SOFIA/HAWC+ observations\cite{Pare2024}.

We describe a study that uses all four PRIMAger bands to observe the polarimetric distribution of the CMZ, as shown in Table \ref{tab:PRIMA}. These bands will probe both the warm and cool dust populations. This data cube of maps can be considered as spectral-polarimetric imaging, which probes the integrated signature of magnetized plasma and dust in the environment to differing depths of the molecular structures. By studying how the magnetic field varies over different temperatures it is therefore possible to disentangle the effects of the magnetic field from changing temperature and density conditions. Furthermore, the PRIMA observations can be compared to observations obtained with other observatories to conduct a multi-scale analysis of the magnetic field. Observations made by PRIMA can also be compared to MHD modeling of the large-scale CMZ field\cite{Tress2024}. The angular resolution of the PRIMAger frequency bands will also be beneficial. The 28$^{\prime\prime}$ beam size of the 235 $\mu$m PRIMA observations can be compared to the similar 19.6$^{\prime\prime}$ SOFIA/HAWC+ observations. This point of comparison can be used to verify the accuracy of the PRIMA observations for high intensity regions.

\subsection{Description of Observations} 
To determine the time required for the PRIMA observations we assume a 1 square degree field of view of the CMZ. We then require a 5$\sigma$ detection of polarized emission of the CMZ assuming the CMZ clouds are 1\% polarized and scaling the minimum Herschel intensities from 70 and 235 $\mu$m\cite{Molinari2011}. To perform the scaling, we assume T = 41 K, $\beta$ = 1.67 for the 70 $\mu$m Herschel observations (warm dust component) and T = 22 K, $\beta$ = 1.67 for the 250 $\mu$m Herschel observations (cool dust component). We scale only the cool dust component to estimate the sensitivity needed for the PRIMA 235 $\mu$m band estimate and scale only the warm dust component to estimate the sensitivity needed for the PRIMA 96 $\mu$m band. For the 126 and 172 $\mu$m band estimates we take the average of the scaling of the warm and cool dust components at those frequencies. This scaling yields the total power sensitivities indicated in Table \ref{tab:PRIMA}. The polarized intensity sensitivities are then obtained by choosing a conservative estimate that the clouds will have 1\% polarization (generally they are polarized at the few \% level). Full polarimetric capability at the shortest wavelength band of PRIMA (96 $\mu$m) is critical to obtain polarimetric observations of the warm dust population. PRIMA is the only proposed next-generation instrument that is designed with full polarimetric capability over these wavelengths -- a unique and valuable feature of this probe. The time required to observe the entire CMZ for all 4 PRIMA bands is shown in the bottom row of Table \ref{tab:PRIMA}. This time is scaled from the 10 hr surface brightness sensitivities shown on the PRIMA webpage\footnote{https://prima.ipac.caltech.edu/page/instruments} for a 1 square degree map.
 
\section{Synergies With Other Observatories} \label{sec:syn}
\begin{table}[ht]
\caption{Comparison with Other Observatories$^i$} 
\label{tab:fonts}
\begin{center}       
\begin{tabular}{|c|c|c|} 
\hline
\rule[-1ex]{0pt}{3.5ex} Observatory & Beam FWHM ($^{\prime\prime}$)  & $\lambda$ ($\mu$m) \\
\hline\hline
\rule[-1ex]{0pt}{3.5ex}  PRIMA & 11, 15, 20, 28 & 96, 126, 172, 235 \\
\hline
\rule[-1ex]{0pt}{3.5ex} SOFIA/HAWC+ & 19 & 214$^{ii}$ \\
\hline
\rule[-1ex]{0pt}{3.5ex}  ALMA & 0.005 - 31 & 315 - 8500 \\
\hline
\rule[-1ex]{0pt}{3.5ex}  LMT & 6, 20 & 1100, 3100 \\
\hline
\rule[-1ex]{0pt}{3.5ex}  FYST & 15, 32, 37, 47, 59 & 350, 730, 850, 1000, 1400 \\
\hline
\rule[-1ex]{0pt}{3.5ex} JCMT (SCUBA-2/POL-2) & 9.6, 14.1 & 450, 850 \\
\hline
\rule[-1ex]{0pt}{3.5ex} PILOT & 132 & 240 \\
\hline
\end{tabular}
\end{center}
\footnotesize{$^i$ Values are taken from the host organization websites which are credited in the acknowledgments at the end of the article.}
\\
\footnotesize{$^{ii}$ Only the 214 $\mu$m SOFIA/HAWC+ band is shown since this is the only band that conducted a CMZ-wide polarimetric study.}
\label{tab:comp}
\end{table}
\begin{figure}
    \begin{center}
    \begin{tabular}{c}
    \includegraphics[width=0.95\textwidth]{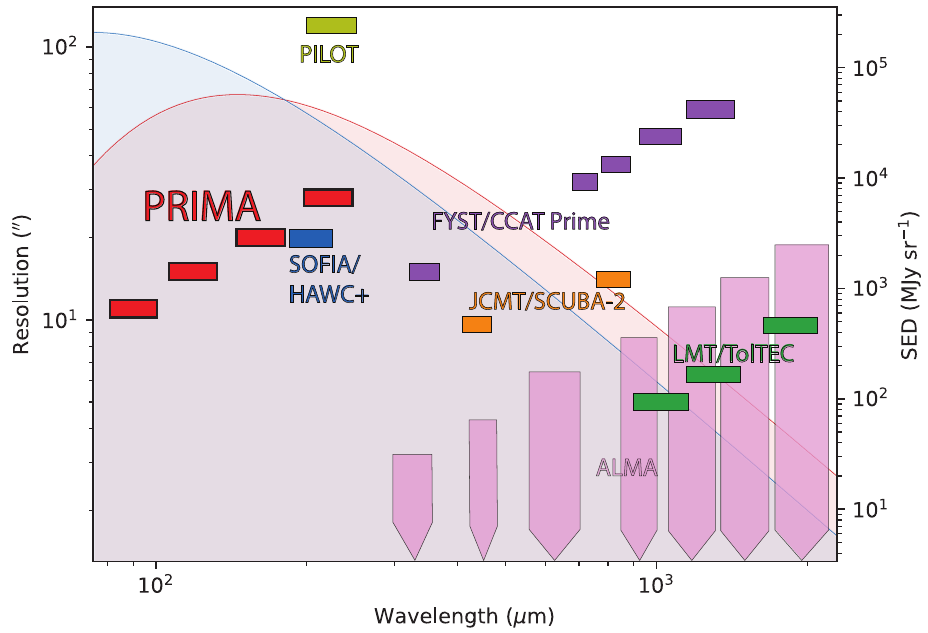}
    \end{tabular}
    \end{center}
    \caption 
    { \label{fig:comp}
The distribution of resolutions and spectral coverage from current and past far-infrared instruments, as listed in Table \ref{tab:comp}. Colored symbols indicate capabilities of different instruments, as labeled. ALMA bands have a range of resolutions as indicated by the shape of the ALMA symbols (Bands 4--10 are shown here). We only show the 214 SOFIA/HAWC+ band because this is the only frequency band that observed the entire CMZ. Greybody curves represent warmer and cooler representative dust SEDs for the Galactic center (blue: T=41 K, $\beta$=1.67; red: T=22 K, $\beta$=1.67). The polarimetric coverage of PRIMA is uniquely positioned for understanding the warmer dust regions in the GC which are heated by energetic events in the CMZ.}
\end{figure}
The properties of the PRIMA observations would synergize with recent, existing, and forthcoming instruments. Table \ref{tab:comp} lists key parameters of the PRIMA observations with current and anticipated capabilities of other instruments. The other instruments we compare to are SOFIA/HAWC+\cite{Harper2018}, PILOT \cite{PILOT2016}, ALMA\cite{Gonzalez2024}, the Large Millimeter Telescope (LMT)\cite{TolTEC2020}, the Fred Young Submillimeter Telescope (FYST)\cite{CCAT2020}, and the JCMT (SCUBA-2/POL-2)\cite{Pol22016}. We note that Figure \ref{fig:comp} and Table \ref{tab:comp} only list the 214 $\mu$m HAWC+ band. This is because only the 214 $\mu$m band observed the entirety of the CMZ in a manner comparable to the PRIMA observations. The observational parameters displayed for ALMA are using the upgraded ALMA2030 specifications\cite{Carpenter2020,Brogan2019}. In this table we compare the resolutions and wavelengths that could possibly be used to observe the warm and cool dust in the CMZ. Other radio instruments, like MeerKAT, that are not able to study the molecular cloud population are not included.

Figure \ref{fig:comp} displays the different telescope resolutions and wavelengths graphically to visualize the unique wavelength and spatial range provided by the PRIMAger imaging bands. The PRIMA bands are shown as red boxes in the figure, and it is clear that PRIMA provides a unique window into the wavelength range of far-infrared emission. 

For example, the observation specifications of PRIMA detailed here would synergize well with ALMA. As discussed previously, the arcsecond and sub-arcsecond resolution capabilities of ALMA would probe protostellar core and envelop spatial scales in the GC, whereas PRIMA is sensitive to molecular cloud scales. The behavior of the magnetic field can be compared at these different scales (in conjunction with MHD modeling) to assess whether the magnetic field or turbulence dominates in CMZ clouds. This CMZ PRIMA project would also synergize well with next-generation radio instruments like the ngVLA and SKA. For example, the ngVLA and SKA will primarily operate at cm wavelengths, a wavelength range that will not be sensitive to the far-infrared dust emission. ngVLA observations can therefore be compared to the $\mu$m PRIMA observations to study connections between the CMZ clouds and the radio non-thermal structures observed in the GC. Similar work has previously been performed using VLA and SOFIA/HAWC+ observations,\cite{Pare2025} but the combined analysis of PRIMA and ngVLA observations will represent a significant improvement in sensitivity to what is possible with current and recent observatories.

These PRIMA observations can also utilize the Wide-field Survey (WFS) that will be conducted using FYST\cite{CCAT2023}. The frequencies of the WFS survey (350, 730, 850, 1000, and 1400 $\mu$m) will complement the goal of studying how the polarization and magnetic field varies as a function of wavelength, enabling more extensive spectro-polarimetric analysis. The PRIMA frequency bands are a vital component of this analysis, making it possible to assess how the polarization and magnetic field varies between the cool and warm dust components in the CMZ as described previously.

We can see from Table \ref{tab:comp} that PRIMA provides unique and crucial capabilities in both resolution and frequency space. These capabilities will be invaluable for furthering our understanding of the role of magnetic fields and the nature of dust in the CMZ. Work of this form will enhance our understanding of Galactic nuclear regions more generally. 

\subsection*{Disclosures}
The authors declare there are no financial interests, commercial affiliations, or other potential conflicts of interest that have influenced the objectivity of this research or the writing of this paper.

\subsection*{Data Availability}
The data used to produce the figures shown in this paper are available on public data archives. The Herschel 250 and 70 $\mu$m observations shown in Figures \ref{fig:legend} and \ref{fig:2-color} are available on the NASA Infrared Science Archive. The SOFIA/HAWC+ observations shown in Figures \ref{fig:2-color} and \ref{fig:CND} are also on the NASA Infrared Science Archive. The data used to create Figure \ref{fig:multi} are available on SIMBAD, CDS, and the ALMA data archive. Observatory parameters listed in Tables \ref{tab:PRIMA} and \ref{tab:comp} are taken from publicly available documentation on the observatory and operating organization websites.
 
\subsection* {Acknowledgments}
SOFIA was jointly operated by the Universities Space Research Association, Inc. (USRA), under NASA contract NNA17BF53C, and the Deutsches SOFIA Institut (DSI) under DLR contract 50 OK 2002 to the University of Stuttgart. Herschel is an ESA space observatory with science instruments provided by European-led Principal Investigator consortia and with important participation
from NASA. The James Clerk Maxwell Telescope is operated by the East Asian Observatory on behalf of The National Astronomical Observatory of Japan; Academia Sinica Institute of Astronomy and Astrophysics; the Korea Astronomy and Space Science Institute; the National Astronomical Research Institute of Thailand; Center for Astronomical Mega-Science (as well as the National Key R\&D Program of China with No. 2017YFA0402700). Additional funding support is provided by the Science and Technology Facilities Council of the United Kingdom and participating universities and organizations in the United Kingdom and Canada. The Caltech Submillimeter Observatory is operated by the California Institute of Technology. The VLA and ALMA are operated by the National Radio Astronomy Observatory, which is a facility of the National Science Foundation operated under cooperative agreement by Associated Universities, Inc. ALMA is a partnership of ESO (representing its member states), NSF (USA) and NINS (Japan), together with NRC (Canada), MOST and ASIAA (Taiwan), and KASI (Republic of Korea), in cooperation with the Republic of Chile. The Joint ALMA Observatory is operated by ESO, AUI/NRAO and NAOJ. The LMT project is a joint effort of the Instituto Nacional de Astrófisica, Óptica, y Electrónica (INAOE) and the University of Massachusetts at Amherst (UMASS).

\bibliography{article}   
\bibliographystyle{spiejour}   


\vspace{2ex}\noindent\textbf{Dylan M. Par\'e} is a postdoctoral scholar at Villanova University. He received his BS degree in physics and astronomy from the University of Massachusetts, Amherst in 2017; his MS in astronomy from the University of Iowa in 2019; and his PhD in physics from the University of Iowa in 2022. His current research interests include cosmic magnetic fields, radio interferometry, and far-infrared dust polarization.

\vspace{2ex}\noindent\textbf{David T. Chuss} is a Professor of Physics at Villanova University. He received his BS in Physics at Villanova University in 1995, an MS in Astronomy and Astrophysics from Penn State in 1997, and a PhD at Northwestern University in 2002.  He was an NRC Postdoc at NASA Goddard and later worked as a Research Astrophysicist there from 2004-2015. He specializes in astronomical polarimetry instrumentation and analysis from far-infrared through millimeter wavelengths. 

\vspace{2ex}\noindent\textbf{Kaitlyn Karpovich} is a PhD student at Stanford University. She received her BS in Physics at Villanova University in 2024. Her research interests include studying interstellar dust and magnetic fields in star-forming regions using far-infrared through millimeter polarimetry. 

\vspace{2ex}\noindent\textbf{Natalie Butterfield} is an assistant scientist at the National Radio Astronomy Observatory. She received her Bachelor degree in physics from the University of Virginia in 2011 and later her Doctorate in Physics from the University of Iowa in 2018. She went on to a postdoctoral position at the Green Bank Observatory (2018-2021) and later at Villanova University (2021-2022). Her current research interests include studying gas dynamics in star forming regions in the inner region of the Milky Way galaxy.

\vspace{2ex}\noindent\textbf{Mark Morris} has been studying the Galactic Center since 1971 using radio, infrared, and X-ray techniques, and has focused much of his work, including polarization measurements, on investigations of the GC magnetic field.

\vspace{2ex}\noindent\textbf{Jeffrey Iuliano} is a visiting assistant professor at Colorado College.  He received his Bachelor degree in physics and philosophy from Harvard University in 2013, and PhD in Astrophysics from Johns Hopkins University in 2020 working on the CLASS CMB survey experiment, particularly focusing on cryogenic receiver development.  He went on to a postdocoral position at the University of Pennsylvania working on the Simons Observatory Large Aperture Telescope, and then at Villanova University working with the FIREPLACE dataset.

\vspace{1ex}
\noindent Biography of Dr. Edward Wollack is not available. Photographs of the authors are not available.

\listoffigures
\listoftables

\end{spacing}
\end{document}